# Rabi Oscillations at Three-Photon Laser Excitation of a Single Rubidium Rydberg Atom in an Optical Dipole Trap

*I.I. Beterov* [a, b, c, d], *E.A. Yakshina* [a, b, d], *G. Suliman* [a, b], *P.I. Betleni* [a, b],
*A.A. Prilutskaya* [a, b], *D.A. Skvortsova* [a, c], *T.R. Zagirov* [a, b], *D.B.Tretyakov* [a, b],
*V.M. Entin* [a], *N.N. Bezuglov* [a, e], *I.I. Ryabtsev* [a, b,] *

[a] *Rzhanov Institute of Semiconductor Physics SB RAS, Pr. Lavretyeva 13, 630090, Novosibirsk, Russia*
[b] *Novosibirsk State University, Ul. Pirogova 2, 630090, Novosibirsk, Russia*
[c] *Novosibirsk State Technical University, Pr. Karla Marksa 20, 630073, Novosibirsk, Russia*
[d] *Institute of Laser Physics SB RAS, Pr. Lavrentyeva 15b, 630090, Novosibirsk, Russia*
[e] *Saint-Petersburg State University, Universitetskaya nab. 7/9, 199034, Saint-Petersburg, Russia*

In an experiment on three-photon laser excitation $5S_{1/2} \rightarrow 5P_{3/2} \rightarrow 6S_{1/2} \rightarrow 37P_{3/2}$ of a single [87]Rb Rydberg atom in an optical dipole trap, we have observed for the first time three-photon Rabi oscillations between the ground and the Rydberg states. The single atom was detected optically by measuring the signal of resonant fluorescence with a low-noise sCMOS video camera. Relative probability of the atom to remain in the trap after the action of three synchronized laser excitation pulses was measured for their durations varied in the range from 100 ns to 2 μs. A specific feature of the experiment was the usage of intense laser radiation at the wavelength of 1367 nm on the second excitation step, which provided the single-photon Rabi frequency up to 2 GHz to control the effective detunings of the intermediate levels of the three-photon transition due to ac Stark effect. We have detected Rabi oscillations with frequency from 1 to 5 MHz depending on the intensities of the laser pulses on the first and the second excitation steps with the coherent time of 0.7-0.8 μs. The ways to increase the coherence time and contrast of the three-photon Rabi oscillations for applications in quantum information processing with Rydberg atoms are discussed.

## 1. Introduction

In the recent years, ultracold neutral atoms became one of the most promising platforms to implement quantum computation. In particular, high-fidelity (>99.5%) two-qubit quantum gates were demonstrated at parallel operation of 60 single Rb atoms [1], and trapping of more than 6000 single Cs atoms in an optical dipole trap array was realized [2]. To perform quantum computation with ultracold neutral atoms, coherent laser excitation of atoms to Rydberg states with the principal quantum number $n \gg 1$ is required. As the orbit radius of Rydberg electron grows as $n^2$, dipole moments of Rydberg atoms also grow as $n^2$, so these atoms interact much stronger than ground-state atoms. Due to large interaction energy of Rydberg atoms, one can obtain quantum entanglement between neutral atoms and implement two-qubit gates [4-6].

As a rule, in experiments on building quantum processors with neutral atoms as qubits, single Rb and Cs atoms are used [6]. For laser excitation of $nS$ and $nD$ Rb Rydberg atoms, the two-photon scheme $5S \rightarrow 5P \rightarrow nS, nD$ is commonly used with 780 nm wavelength on the first step and 480 nm wavelength on the second step [7,8], or another two-photon scheme $5S \rightarrow 6P \rightarrow nS, nD$ is used with 420 nm wavelength on the first step and 1013 nm wavelength on the second step. In experiments with single Cs atoms, the two-photon scheme $6S \rightarrow 7P \rightarrow nS, nD$ is typically used with 459 nm wavelength on the first step and 1040 nm wavelength on the second step [9]. One-photon excitation of Rydberg states is also possible with ultraviolet laser radiation (297 nm wavelength for Rb atoms [10]). Two-photon excitation with counterpropagating laser beams partly suppresses the residual Doppler effect that appears due to finite atom temperature in optical traps leading to the loss of coherence [11]. Unlike to two-photon and three-photon schemes, one-photon excitation does not provide even partial compensation of the Doppler effect.

Three-photon laser excitation of single Rydberg atoms has a number of advantages. First, it allows excitation of $nP$ Rydberg states, for which novel schemes of three-qubit Toffoli gates were proposed [12,13]. Unlike to atoms in the Rydberg $nS$ and $nD$ gates, single $nP$ alkali-metal Rydberg atoms and their interactions in fact are poorly investigated experimentally.

Second, three-photon laser excitation makes it possible to almost completely suppress the recoil effect and the Doppler effect with a star-like geometry of three laser beams with a zero sum of their wave vectors [14]. This holds promise for increasing the fidelity of quantum gates.

Third, in the case of two-photon laser excitation, due to the large decay rate of intermediate excited states, it is necessary to introduce large detunings from the exact resonance with the intermediate states (>1 GHz). Our theoretical analysis showed that three-photon laser excitation with a high Rabi frequency at the second step compared to the Rabi frequencies at the first and third steps allows achieving the coherent excitation regime even when all three steps are tuned to the exact resonance with the atomic transitions. In this case, the absence of population and spontaneous decay of the intermediate states is ensured by light shifts due to the strong radiation on the second step.

In previous works [15-18] we experimentally investigated the spectra and dynamics of three-photon laser excitation $5S_{1/2} \rightarrow 5P_{3/2} \rightarrow 6S_{1/2} \rightarrow nP$ of cold Rb Rydberg atoms in a working magneto-optical trap using cw single-frequency lasers at each step and detecting single Rydberg atoms using the selective field ionization method [3].

In our last paper [19], three-photon laser excitation of a single $^{87}$Rb atom in an optical dipole trap into the $37P_{3/2}$ Rydberg state by laser radiation with wavelengths of 780 nm, 1367 nm, and 743 nm was demonstrated experimentally for the first time using the scheme $5S_{1/2} \rightarrow 5P_{3/2} \rightarrow 6S_{1/2} \rightarrow 37P_{3/2}$. Excitation into Rydberg states was detected optically by the losses of atoms in the optical dipole trap. The spectra of pulsed three-photon laser excitation of a single Rydberg atom were recorded. The spectrum width was 2 MHz. The dependence of the excitation probability on the laser pulse duration was also measured, but no signatures of Rabi population oscillations between the ground and Rydberg states were observed. Such oscillations are necessary for the subsequent implementation of quantum gates with Rydberg atoms.

In this paper, we present the results of our new experiment on three-photon laser excitation of a single $^{87}$Rb Rydberg atom trapped in an optical dipole trap. By narrowing the line widths of the second and third step lasers, we were able to observe for the first time three-photon Rabi population oscillations with frequencies from 1 to 5 MHz depending on the intensities of the laser pulses of the first and second excitation steps. A specific feature of the experiment was the use of intense laser radiation with 1367 nm wavelength at the second excitation step, providing a single-photon Rabi frequency of up to 2 GHz to control the effective detunings of the intermediate levels of the three-photon transition due to the dynamic Stark effect. We also discuss the ways to increase the coherence time and contrast of three-photon Rabi oscillations for applications in quantum information processing with Rydberg atoms.

## 2. Experimental setup

The experimental setup, described in detail in our previous paper [19] on three-photon laser excitation of a single $^{87}$Rb Rydberg atom, is shown in Fig. 1. The main changes in the present work include narrowing the spectral width of the lasers, optimizing the laser beam intensities, and upgrading the optical schemes for modulating the laser radiation.

*Capture of atoms in an optical dipole trap.* [87]Rb atoms are cooled and captured in a magneto-optical trap (MOT) in a vacuum chamber, in the center of which a cloud of cold atoms with a temperature of 80-100 μK is formed. Then, to capture atoms from the MOT into the optical dipole trap, radiation from a laser system with a wavelength of 850 nm is used, based on the Eagleyard EYP-DFB-0852 master DFB laser and a Toptica Boosta Pro semiconductor amplifier with an output power of 1.4 W. It is modulated using an acousto-optic modulator (AOM) and then fed into the optical system via optical fiber. After exiting the optical fiber, the dipole trap laser radiation is collimated, reflected from a dichroic mirror, passed through a polarization beam splitter and then focused into a cloud of cold rubidium atoms by an objective with a focal length of *f*=119 mm and a numerical aperture of NA=0.172. An expanding telescope consisting of two lenses with focal lengths of *f*=75 mm and *f*=500 mm is installed in front of the objective. The radiation of the optical dipole trap is focused into a spot with a diameter of 8-9 μm at an intensity level of e[-2].

*Detection of trapped atoms.* To detect the trapped [87]Rb atoms, resonance fluorescence induced by cooling lasers with a wavelength of 780 nm (not shown in Fig. 1) is used. Spontaneously emitted photons are collected by the same objective with a focal length of *f* = 119 mm, passed through the telescope, are partially reflected from a polarization beam splitter, and then are focused by a lens of *f* = 50 mm onto a digital sCMOS video camera Tucsen Dhyana 400D. Two interference filters are installed in front of the video camera, transmitting radiation only at a wavelength of 780 nm.

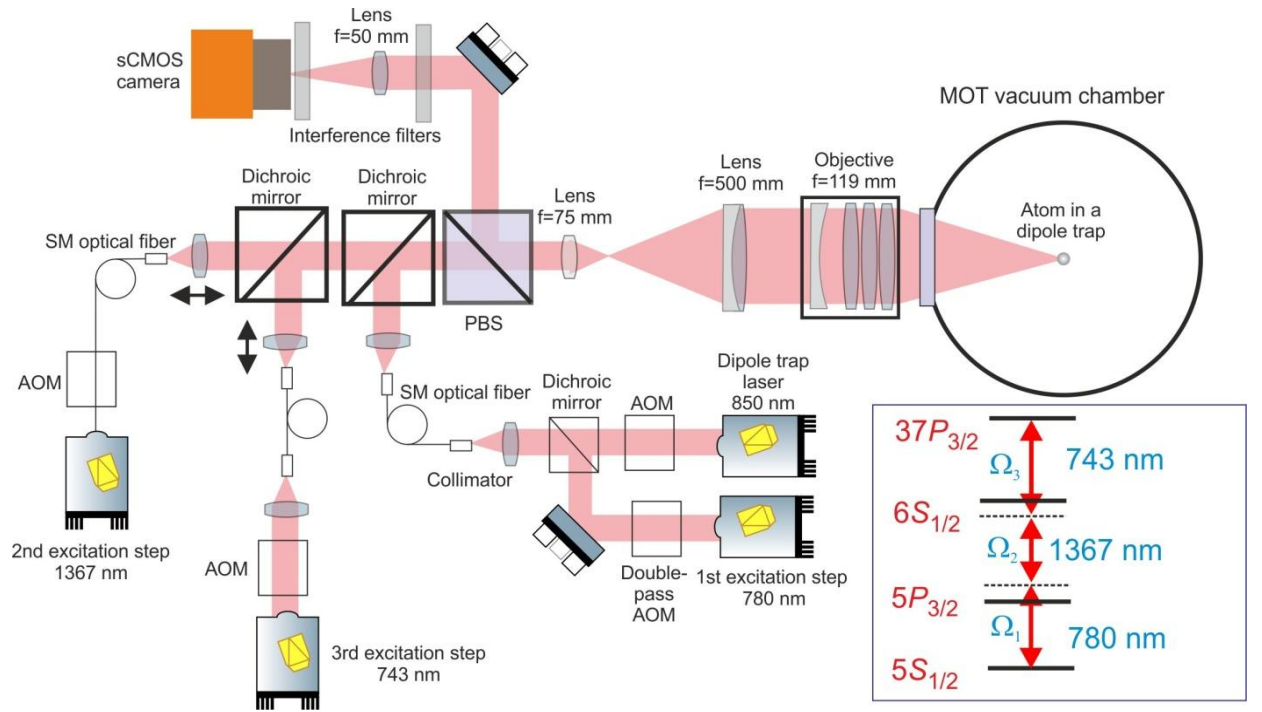

**Fig. 1.** Experimental setup for implementing coherent three-photon laser excitation to Rydberg states with the scheme $5S_{1/2} \rightarrow 5P_{3/2} \rightarrow 6S_{1/2} \rightarrow 37P_{3/2}$ (shown in the inset) of a single [87]Rb atom trapped in an optical dipole trap. Full description is given in the text.

*Laser excitation of Rydberg states.* The first step of laser excitation uses a Toptica DL Pro external-cavity diode laser with a Toptica Boosta Pro semiconductor amplifier. The radiation of this laser is used for both laser cooling and laser excitation of atoms into Rydberg states. The first-step laser frequency is locked to the saturated absorption resonances in [87]Rb atoms (cross-resonance between the hyperfine sublevels $|F=2\rangle$ and $|F'=3\rangle$ of the $5P_{3/2}$ state. AOMs are used to independently control the frequency detunings of the cooling radiation and the first-step laser excitation radiation. The first-step laser radiation frequency was shifted using an AOM operating in a double-pass configuration so that the radiation had a blue detuning $\delta_1$=+30 MHz from the resonance $|F{=}2\rangle{\to}|F'{=}3\rangle$, as shown in the inset in Fig. 1. The first-step laser excitation radiation was combined with the radiation of a dipole trap in front of the optical fiber using a dichroic mirror.

The second step of laser excitation uses a Sacher Lasertechnik external-cavity diode laser with a wavelength of 1367 nm. The laser radiation frequency is locked by the Pound-Drever-Hall method to one of the transmission peaks of a highly-stable Fabry-Perot interferometer by Stable Lasers using the Vescent Photonics D2-125PL locking system. As a result, the radiation has a red frequency detuning from the resonance with the $5P_{3/2}$ ($F{=}3$)$\to6S_{1/2}$($F{=}2$) transition $\delta_2$=-60 MHz, as shown in the inset in Fig. 1. The estimated line width from the error signal is less than 25 kHz, and from the beat spectrum obtained by the self-heterodyne mixing method, it is less than 5 kHz. An AOM is used for amplitude modulation of the second-step laser radiation.

The third step uses a single-frequency Ti:Sapphire ring resonator laser from Tekhnoscan pumped by a Coherent Verdi G5 solid-state Nd:YVO$_4$ laser. The laser frequency is stabilized by the transmission peak of a highly-stable Fabry-Perot interferometer from Stable Laser Systems using the Pound-Drever-Hall error signal generation system and a Vescent Photonics D2-125 PID controller. In order to ensure the ability to tune to an arbitrary Rydberg state, the frequency is locked by sideband components that arise when mixing an radio-frequency (RF) signal with an arbitrarily specified frequency in the range from 10 MHz to 200 MHz to the input of the electro-optical modulator in the frequency stabilization system. The RF signal is synthesized using a Rigol DG4202 generator controlled via a LAN interface. The third-step laser line width is estimated by the error signal to be no more than 2 kHz. An AOM is installed for amplitude modulation of the third-step laser. The radiation frequency is controlled by a precision wavelength meter WS-U from Angstrom.

The radiation of the second-step laser with a wavelength of 1367 nm and a power of up to 1 mW and the third-step laser with a wavelength of 743 nm and a power of 10-50 mW are fed into the optical system using separate optical fibers, as shown in Fig. 1. The laser beams are combined on a dichroic mirror and fed into the optical system coaxially with the beam of the optical dipole trap. Using lenses mounted on precision movable mounts that collimate the radiation at the output of the optical fibers, the divergence of the second- and third-step laser beams is individually adjusted so that their focal spots with a diameter of no more than 10 μm at the intensity level e$^{-2}$ exactly coincide with the focal spot of the optical dipole trap. For this purpose, the DataRay BeamMap2 laser radiation intensity profile meter is used. All beams from the exciting lasers pass through a polarization beam splitter and have horizontal polarization, like the radiation of the optical dipole trap laser.

*Timing diagram of the experiment*. The timing diagram of the experiment is shown in Fig. 2. The experimental setup is controlled by a SpinCore PulseBluster programmable timer board. [87]Rb atoms are initially loaded into the MOT for 0.1-5 seconds, and the optical dipole trap is loaded simultaneously. The dipole trap laser is modulated by square pulses with a frequency of 1 MHz and a duty cycle of 60% to avoid the influence of light shifts on the detection of atoms in the absence of the trap laser. A Tucsen Dhyana 400D digital sCMOS video camera detects atoms as a sequence of images with an exposure time of 175 ms until the moment of loading single atoms into the trap and the appearance of the first resonance fluorescence signal from the trapped atoms.

After the detection of a single atom, the procedure of laser excitation of atoms into Rydberg states and optical detection of Rydberg excitation is started. The cooling lasers and the gradient magnetic field of the MOT are switched off. Then the beams of the cooling laser, the pumping laser and the video camera are switched on for the first detection of fluorescence signal from the trapped atom to confirm that the atom is held in the optical dipole trap. After that, the cooling laser is switched off, and the repumping laser remains on for 2 ms. This ensures optical pumping of the trapped atom into the state $5S_{1/2}$($F{=}2$). After that, the radiation of the optical dipole trap is switched off to eliminate the light shifts associated with this radiation, and all three steps of laser excitation of Rydberg atoms are switched on. After a time of 0.1-5 μs, the radiation of the exciting lasers is switched off, and the radiation of the optical dipole trap is switched on again. The timing diagram of the laser excitation pulses is discussed in detail in Section 4.

The intense radiation of the dipole trap laser blows the $^{87}$Rb atom away from the dipole trap if it is in the Rydberg state, and recaptures it if it is in the ground state. Then the beams of the cooling laser, the repumping laser, and the video camera are turned on again. The atom that was not excited to the Rydberg state and remained in the optical dipole trap is detected again. In the experiment with single atoms, the probability of re-detection of the atom is measured as dependence on the frequency of the laser radiation.

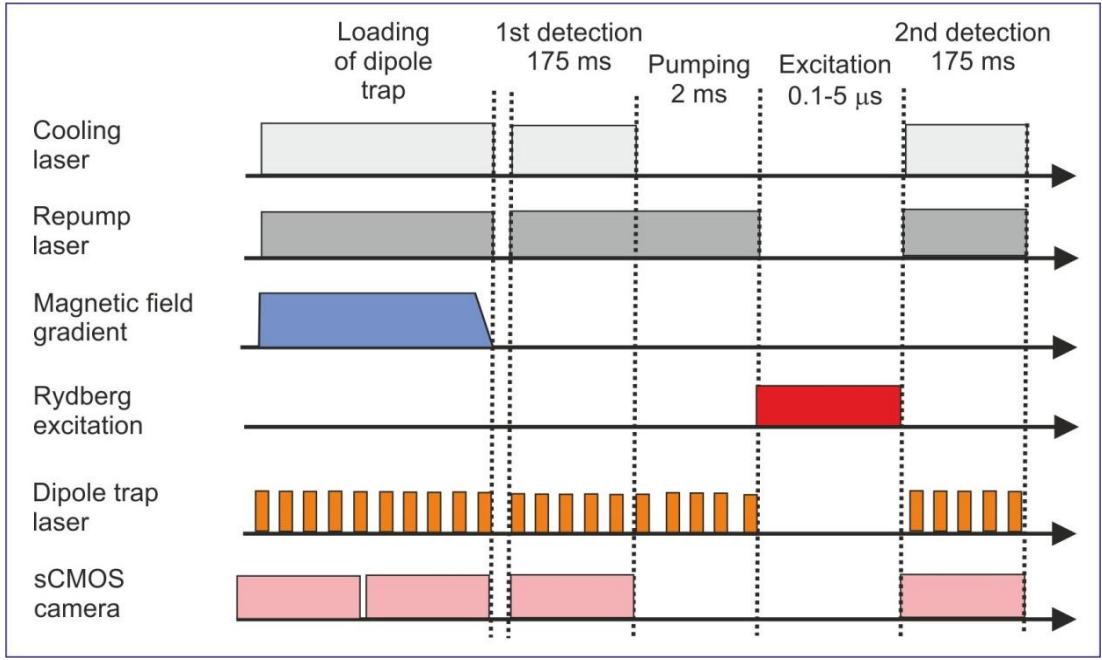

**Fig. 2.** Timing diagram of the experiment on coherent three-photon laser excitation of a single $^{87}$Rb atom to Rydberg state. Full description is given in the text.

### 3. Theory of three-photon laser excitation of Rydberg state in a single atom

In the theoretical description we denote the ground state $5S_{1/2}(F=2)$ as state 1, the first intermediate state $5P_{3/2}(F=3)$ with a radiative lifetime of 27 ns as state 2, the second intermediate state $6S_{1/2}(F=2)$ with a radiative lifetime of 50 ns as state 3, and the Rydberg state $37P_{3/2}$ with a radiative lifetime of 43 μs as state 4 (the diagram of levels and transitions is shown in the inset of Fig. 1). For each intermediate single-photon transition $j = 1, 2, 3$ we introduce the corresponding Rabi frequency $\Omega_j = d_j E_j / \hbar$ (here $d_j$ are the dipole moments of single-photon transitions, $E_j$ are the amplitudes of the electric field for linearly polarized light fields) and the detuning $\delta_j$. Scanning the full detuning $\delta = \delta_1 + \delta_2 + \delta_3$ of the three-photon transition can be performed by scanning the frequency of any laser. In our experiments, the detuning of the third-step laser $\delta_3$ is scanned.

As we have shown in Ref. [14], in the absence of spontaneous relaxation of all levels and at sufficiently large detunings from intermediate resonances $\Omega_1 \ll |\delta_1|, \Omega_2 \ll |\delta_2|$, the population of the Rydberg state can be calculated by solving the Schrödinger equation to find the

probability amplitudes $a_j$ of each level $j = 1\text{-}4$ in the rotating wave approximation. As a result, we obtain the following dependence of population of the Rydberg state on the interaction time $t$ for fully coherent three-photon laser excitation:

$$|a_4|^2 \approx \frac{\Omega^2}{\Omega^2 + (\delta + \Delta_1 + \Delta_4)^2}\left[1 - \cos\left(t\sqrt{\Omega^2 + (\delta + \Delta_1 + \Delta_4)^2}\right)\right]\bigg/2, \tag{1}$$

where $\Omega = \Omega_1\Omega_2\Omega_3/(4\delta_1\delta_3)$ is the effective Rabi frequency for three-photon excitation, and $\Delta_1 = \Omega_1^2/(4\delta_1)$ and $\Delta_4 = \Omega_3^2/(4\delta_3)$ are the light shifts of states 1 and 4, respectively. Equation (1) shows that the condition for exact three-photon resonance is $\delta + \Delta_1 + \Delta_4 = 0$, with the population oscillating between the ground and Rydberg states at frequency $\Omega$. Equation (1) also describes the excitation spectrum of the Rydberg state when scanning $\delta$ for a fixed interaction time $t$.

In Eq. (1), the Rabi oscillations last infinitely long. In a more realistic theoretical model that takes into account the spontaneous relaxation of excited levels 2–4, the Rabi oscillations decay with a decay constant $\gamma = 1/\tau$ determined by the inverse lifetime of the Rydberg state $\tau$, or even faster if the detunings of the intermediate resonances are not large enough. In this case, the population of the Rydberg state reaches a certain stationary value. We have previously constructed such a model in a four-level approximation based on the optical Bloch equations for the density matrix [15]. It is impossible to find an exact analytical solution for the population $\rho_{44}$ of the Rydberg state at arbitrary intermediate Rabi frequencies and detunings, so in the general case it is necessary to solve the problem numerically. However, for large detunings of the laser radiation frequencies from the intermediate resonances, our four-level system can be approximated by an effective two-level system with a direct optical transition $1 \rightarrow 4$.

For a two-level system in Ref. [15] we have found the following approximate analytical solution in the case of strong ($\Omega \gg \gamma$) excitation:

$$\begin{aligned}
\rho_{44} \approx &\frac{\Omega^2}{2\Omega^2 + \gamma^2 + 4\delta^2}\left[1 - e^{-\frac{2\Omega^2 + \delta^2}{4\Omega^2 + \delta^2}\gamma t}\right] + \\
&\frac{\Omega^2/2}{\Omega^2 + \delta^2}\left[e^{-\frac{2\Omega^2 + \delta^2}{4\Omega^2 + \delta^2}\gamma t} - e^{-\frac{6\Omega^2 + \delta^2}{4\Omega^2 + \delta^2}\gamma t/2}\cos\left(t\sqrt{\Omega^2 + \delta^2}\right)\right]
\end{aligned} \tag{2}$$

Comparison of Eq. (2) with the simpler Eq. (1) for the model without relaxation shows that Eq. (1) is applicable at short interaction times ($\gamma t \ll 1$). At long interaction times ($\gamma t \gg 1$), the Rabi oscillations decay and the population reaches a stationary value described by the Lorentzian contour of the excitation spectrum. Comparison of Eq. (2) with the results of an exact numerical calculation showed that already at $\Omega > 3\gamma$ it has good accuracy. In order to apply Eq. (2) to the analytical description of the spectrum of coherent three-photon excitation, it is necessary to make the following replacements in Eq. (2): $\Omega = \Omega_1\Omega_2\Omega_3/(4\delta_1\delta_3)$ is the three-photon Rabi frequency, $\gamma$ is the inverse lifetime of the Rydberg state, and make a substitution $\delta \rightarrow \delta + \Delta_1 + \Delta_4$ to take into account the light shift of the three-photon resonance.

At exact resonance ($\delta = 0$), Eq. (2) gives the maximum amplitude of Rabi oscillations and has the following dependence on time:

$$\rho_{44} \approx \frac{\Omega^2}{2\Omega^2 + \gamma^2}\left[1 - e^{-\gamma t/2}\right] + \frac{1}{2}\left[e^{-\gamma t/2} - e^{-3\gamma t/4}\cos(\Omega t)\right]. \tag{3}$$

Since the condition $\Omega >> \gamma$ must be satisfied, Eq. (3) can be simplified and the following dependence can be obtained for approximating experimental Rabi oscillations:

$$\rho_{44} \approx \frac{1}{2}\left[1 - e^{-3\gamma t/4}\cos(\Omega t)\right]. \qquad (4)$$

In addition, the developed theoretical model [15] allows one to phenomenologically take into account the finite line widths $\Gamma_i$ of all three lasers in the phase diffusion model, when random phase fluctuations are present in the laser radiation, but amplitude fluctuations are absent [20]. To take them into account, an imaginary part equal in modulus to $\Gamma_i / 2$ is added to each detuning $\delta_i$ in the density matrix equations for optical coherences to introduce additional decay in the coherence. Our numerical calculations according to this model showed good agreement between experiment and theory [15-18]. However, it should be borne in mind that in such a model the laser radiation spectrum has a Lorentzian shape, while lasers usually have a Gaussian profile with a faster decay on the wings. Therefore, the theoretical excitation spectra of Rydberg states in this model may have some discrepancies with the experimental data on the resonance wings.

## 4. Methodology for conducting experiments on coherent three-photon laser excitation of a single $^{87}$Rb atom into a Rydberg state

The time diagram of the laser pulses of steps 1–3 in the experiment on coherent three-photon laser excitation of a single $^{87}$Rb atom into the Rydberg state is shown in Fig. 3. The laser pulse of step 1 (wavelength 780 nm) was formed using an AOM operating in a double-pass scheme with a focused beam and had the shortest edges (about 50 ns). The laser pulses at steps 2 and 3 (1367 nm and 743 nm) were formed using AOMs in single-pass schemes without beam focusing and had edge durations of about 100 ns and 200 ns, respectively. Therefore, to ensure the best time resolution and contrast when recording Rabi oscillations, the following time sequence was adopted, which is the main feature of the three-photon laser excitation of the Rydberg state.

The first pulse to be switched on was the step 2 laser pulse. Its intensity was chosen to be high enough to provide a single-photon Rabi frequency at the transition $5P_{3/2}(F=3) \rightarrow 6S_{1/2}(F=2)$ of about 2 GHz. Since the radiation of this laser had a red detuning of $\delta_2 = -60$ MHz from the exact transition frequency, it created large light shifts of two levels (~1 GHz), with one level $5P_{3/2}(F=3)$ shifted down in energy and the other $6S_{1/2}(F=2)$ up. Thus, the intense radiation of step 2 formed the so-called "dressed" quasi-energy levels, which then provided large intermediate detunings for the radiation of step 1 and step 3 lasers from their resonances with atoms.

The second to be switched on was the step 3 laser with a delay of 0.5 μs after switching on the step 2 laser. Initially, this radiation had a small detuning of $\delta_3 = +30$ MHz from the exact $6S_{1/2}(F=2) \rightarrow 37P_{3/2}$ transition frequency, but due to the light shift of the state $6S_{1/2}(F=2)$ from the previously switched on radiation of laser 2, the frequency detuning of the step 3 laser changed to a value of $\delta_3 \sim -1$ GHz. Due to such large detunings, in the absence of the radiation of the step 1 laser, the populations of all atomic levels did not change in the field of the switched on radiation of the steps 2 and 3 lasers.

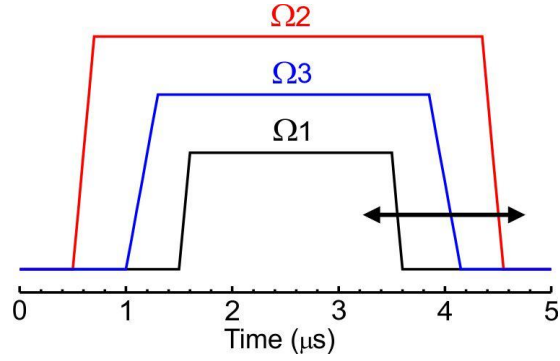

**Fig. 3.** Timing diagram of the laser pulses on the 1-3 steps in the experiment on coherent three-photon laser excitation of a single [87]Rb atom to Rydberg state. Full description is given in the text.

The laser of step 1 was switched on last, with a delay of 0.5 μs after switching on the laser of step 3. Initially, this radiation also had a small detuning of $\delta_1 = +30$ MHz from the exact $5S_{1/2}(F=2) \rightarrow 5P_{3/2}(F=3)$ transition frequency, but due to the light shift of the state $5P_{3/2}(F=3)$ from the previously switched on radiation of laser 2, the frequency detuning of laser step 1 changed to a value of $\delta_1 \sim +1$ GHz.

With the switching on of the first-step laser, three-photon excitation of the Rydberg state began according to the scheme $5S_{1/2} \rightarrow 5P_{3/2} \rightarrow 6S_{1/2} \rightarrow 37P_{3/2}$. Despite the large light shifts of the intermediate levels under the action of the intense radiation of the second-step laser, the three-photon resonance arose near the detuning of the third-stage laser $\delta_3 = +30$ MHz from the exact $6S_{1/2}(F=2) \rightarrow 37P_{3/2}$ transition frequency, since the condition of the exact three-photon resonance $\delta = \delta_1 + \delta_2 + \delta_3 \approx 0$ was fulfilled for this detuning. In general, the values of small initial detunings from the exact atomic resonances practically did not affect the observation of Rabi oscillations, since the detunings of the intermediate levels during three-photon excitation were determined by their huge (~1 GHz) light shifts under the action of the intense radiation of the second-step laser. This is a remarkable feature of three-photon excitation that we predicted during the preliminary theoretical analysis and implemented in this experiment.

The duration of the three-photon excitation pulse was set by the duration of the laser pulse of step 1. Its duration could be set from 100 ns to 5 μs, while at short times it was necessary to take into account the effective duration for the 50 ns edges, i.e. to recalculate for the pulse area.

The three laser pulses were switched off in the reverse order. The pulse of step 1 was switched off first, 0.5 μs after it - the pulse of step 3, then after another 1 μs - the pulse of step 2. Due to the longer retention of the intense pulse of step 2, large detunings of the intermediate single-photon transitions from the frequencies of the lasers of steps 1 and 3 were ensured. For step 3, this made it possible to avoid parasitic reverse population repumping during single-photon de-excitation of the Rydberg state according to the scheme $37P_{3/2} \rightarrow 6S_{1/2}(F=2)$ with subsequent fast spontaneous decay of the state $6S_{1/2}$ through the intermediate state $5P$ into the ground state $5S_{1/2}$. When varying the duration of the pulse of step 1, the pulses of steps 2 and 3 were also synchronously varied so that the time sequence of pulse switching off remained unchanged.

One microsecond after the laser pulse of step 2 was switched off, the dipole trap laser radiation was switched on, and then the presence of a single [87]Rb atom was detected by its resonance fluorescence signal. If the atom was in the Rydberg state, the trap laser radiation quickly (in a few microseconds) blew it away from the trap, and the fluorescence signal was absent. If the atom was in the ground state, it was recaptured in the trap, and the fluorescence signal was present. In this way, the population $\rho_{11}$ of a single atom in the ground state $5S_{1/2}$ was measured after the end of the action of the laser pulses of three-photon excitation. Since the intermediate levels of the three-photon transition are practically not populated due to their large detunings from resonances with all laser radiations, the experimentally measured population (excitation probability) of the Rydberg state can be further determined with good accuracy as $\rho_{44} \approx 1 - \rho_{11}$.

In the experimental records presented below, each point was averaged over 100 measurements. Since each measurement could be accompanied by the loss of an atom when it was excited to the Rydberg state, it could take several seconds to recapture a single atom, and the total time required for one record was about half an hour. Therefore, slow drifts in the frequency and power of the laser radiation could affect the results obtained.

## 5. Experimental results and comparison with theory

Figure 4(a) presents an experimental record (blue dots) of the spectrum of three-photon laser excitation $5S_{1/2} \to 5P_{3/2} \to 6S_{1/2} \to 37P_{3/2}$ of the Rydberg state $37P_{3/2}$ in a single $^{87}$Rb atom at an interaction time (laser pulse duration of step 1) of 0.4 μs. The frequency detuning $\delta_3$ of step 3 laser from the exact $6S_{1/2}(F=2) \to 37P_{3/2}$ transition frequency was scanned. A contour close to Lorentzian with a width at half-maximum of 3 MHz was observed, and the resonance amplitude at the center reached $\rho_{44} \approx 1 - \rho_{11} \approx 0.7$.

Then the frequency of step 3 laser was tuned to the center of three-photon resonance, and the dependence of $\rho_{11}$ on the laser pulse duration on step 1 was measured. The obtained dependence (blue dots) is shown in Fig. 1(b). It clearly demonstrates the presence of Rabi oscillations. When approximating this dependence with the analytical Eq. (4), it was determined that the frequency of Rabi oscillations was $\Omega/(2\pi) \approx 1.35$ MHz, and their coherence time (attenuation at the level of $e^{-1}$) was $\tau_{coher} \approx 0.8$ μs. In this case, the first oscillation with amplitude $\rho_{44} \approx 1 - \rho_{11} \approx 0.7$ occurs at a duration of 0.4 μs, which exactly corresponds to the spectrum in Fig. 4(a).

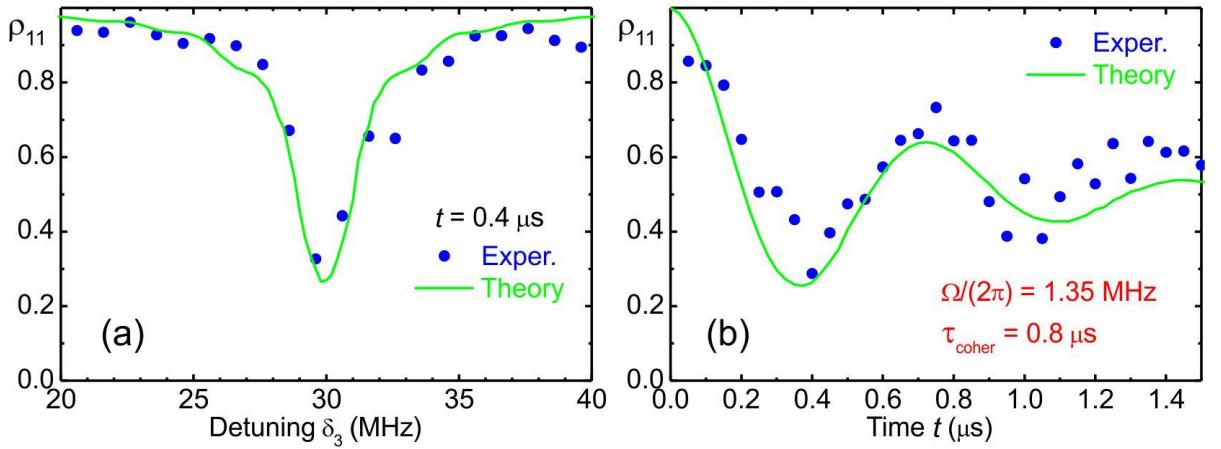

**Fig. 4.** (a) Experimental record (blue dots) of the spectrum of three-photon laser excitation $5S_{1/2} \to 5P_{3/2} \to 6S_{1/2} \to 37P_{3/2}$ of the Rydberg state $37P_{3/2}$ for the interaction time 0.4 μs (duration of the first-step laser pulse). Detuning $\delta_3$ of the third-step laser frequency from the exact $6S_{1/2}(F=2) \to 37P_{3/2}$ transition frequency is scanned. (b) Experimental record (blue dots) of the Rabi oscillations at frequency 1.35 MHz in the center of three-photon resonance. Green solid curves are the results of numerical modeling in a foul-level system with relaxation at one-photon Rabi frequencies $\Omega_1/(2\pi) \approx 80$ MHz, $\Omega_2/(2\pi) \approx 1740$ MHz, $\Omega_3/(2\pi) \approx 30$ MHz.

The previously measured widths of the laser radiation spectra at each step were $\Gamma_1 \approx$ 100 kHz, $\Gamma_2 \approx 5$ kHz, $\Gamma_3 \approx 2$ kHz. Thus, the total line width of the three lasers was $\Gamma \approx 107$ kHz. Therefore, the expected coherence time should have been 9 μs, which is an order of magnitude greater than in the experiment. Also, the amplitude of the first oscillation should have been close to 0.95, which is significantly greater than in the experiment. It was concluded that in the three-photon resonance there is an additional parasitic broadening of ~1 MHz, which is responsible for a significant decrease in the coherence time and the amplitude of the Rabi oscillations.

As a rule, parasitic broadenings of resonances in experiments with Rydberg atoms are caused by parasitic electric fields [17], since the polarizabilities of Rydberg states grow as $n^7$ with increasing principal quantum number $n$. However, in our experiment, a single $^{87}$Rb atom was located in the center of the vacuum chamber and far from all surfaces (~10 cm) on which parasitic electric charges might be present. Therefore, the influence of parasitic electric fields from the walls or windows of the vacuum chamber was practically excluded.

The only source of parasitic broadening could be the not fully compensated laboratory magnetic field in the center of the vacuum chamber. Although it is compensated along three axes by the existing compensating Helmholtz coils during the MOT adjustment, the accuracy of its compensation is 50-100 mG. Such a residual magnetic field leads to Zeeman splitting and broadening of each of the optical transitions between the used $S$ and $P$ levels by up to 250-500 kHz.

In Fig. 4, the green solid lines show the results of numerical modeling with the introduction of a parasitic Zeeman broadening of $\Gamma_z \approx 300$ kHz at each step of laser excitation into the four-level model. Only with the introduction of such broadening it was possible to simultaneously obtain good agreement between experiment and theory for both the spectrum and the Rabi oscillations.

The fitting parameters of the theoretical model were single-photon Rabi frequencies $\Omega_1/(2\pi) \approx 80$ MHz, $\Omega_2/(2\pi) \approx 1740$ MHz, $\Omega_3/(2\pi) \approx 30$ MHz, while the detunings and line widths of the lasers were taken equal to those measured experimentally. For one-photon Rabi frequencies, there was some uncertainty in measuring the intensities of laser pulses, since the real diameters of the waists of focused laser beams in the region of the trapped atom are known with finite accuracy. The powers of laser radiation of steps 1-3 were equal to 150 μW, 180 μW and 3 mW, respectively. The measured diameters of the waists of the lasers of the second and third steps were 10 μm, and the diameter of the waist of the first step laser in the region where the atom is located was 200 μm. Based on this, estimates were obtained for single-photon Rabi frequencies in a theoretical model, taking into account the known (for transitions of steps 1 and 2) and calculated (for step 3) matrix elements of dipole moments.

The results of this experiment led to the conclusion that the parasitic Zeeman broadening limits the contrast and the number of observed Rabi oscillations. To improve them to the theoretical limit, it is necessary to either better compensate for the residual magnetic field, or impose a sufficiently large external magnetic field and implement transitions between certain Zeeman sublevels of the ground and Rydberg states, or simply increase the laser radiation intensity and achieve a significant increase in the Rabi frequency with the same coherence time. The first two options require additional research and will be implemented by us in the near future. In this paper, we used the last option.

It should be noted that to increase the three-photon Rabi frequency, one should increase the radiation intensity of either step 1 or step 3. Increasing the intensity of step 2, on the contrary, will lead to a decrease in the three-photon Rabi frequency due to an increase in the light shifts of intermediate levels and detuning of laser radiation from single-photon resonances.

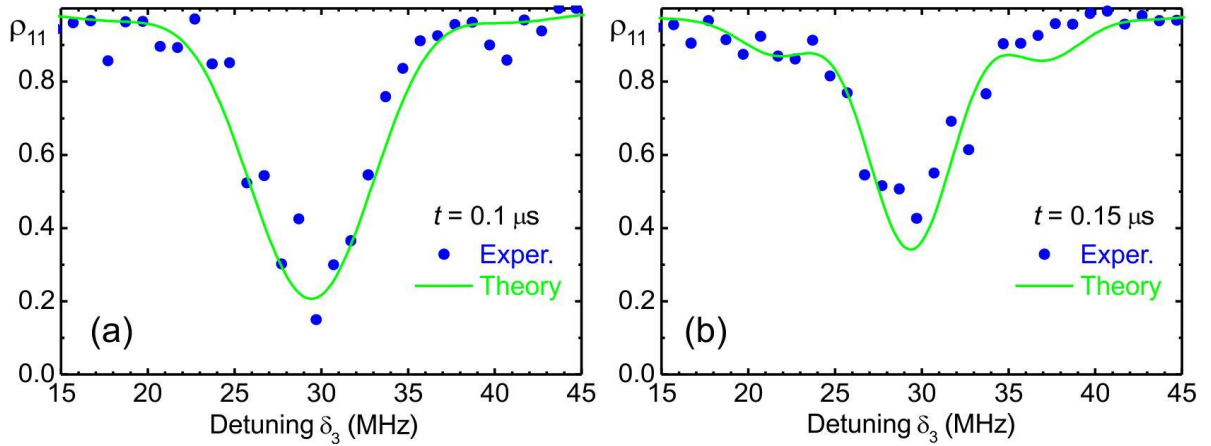

**Fig. 5.** Experimental records (blue dots) of the spectra of three-photon laser excitation $5S_{1/2} \rightarrow 5P_{3/2} \rightarrow 6S_{1/2} \rightarrow 37P_{3/2}$ of the Rydberg state $37P_{3/2}$ at increased by ~10 times (compared to Fig. 4) power of radiation of the first-step laser and interaction time (a) 100 ns and (b) 150 ns. Detuning $\delta_3$ of the third-step laser frequency from the exact $6S_{1/2}(\mathrm{F}=2) \rightarrow 37P_{3/2}$ transition frequency is scanned. Green solid curves are the results of numerical modeling in a foul-level system with relaxation at one-photon Rabi frequencies $\Omega_1/(2\pi) \approx 260$ MHz, $\Omega_2/(2\pi) \approx 1740$ MHz, $\Omega_3/(2\pi) \approx 30$ MHz. The model also takes into account the durations of the 50 ns edges of the laser pulse.

Therefore, in the next experiment, the laser power at step 1 was increased by ~10 times, due to which the single-photon Rabi frequency at step 1 increased from 80 to 260 MHz. Figure 5 shows experimental records (blue dots) of the spectra of three-photon laser excitation $5S_{1/2} \rightarrow 5P_{3/2} \rightarrow 6S_{1/2} \rightarrow 37P_{3/2}$ of the $37P_{3/2}$ Rydberg state with a laser power at step 1 increased by ~10 times (compared to Fig. 4) and an interaction time of (a) 100 ns and (b) 150 ns.

From Fig. 5(a) it can be seen that at an interaction time of 100 ns the resonance amplitude increased to $\rho_{44} \approx 1 - \rho_{11} \approx 0.8$ and the width to 6.5 MHz. The latter is due to the increase in the Fourier width of the short laser pulse. With an increase in the duration to 150 ns in Fig. 5(b) the resonance amplitude decreased to 0.6 and the width to 5 MHz. When numerically modeling these spectra, it was found that for ideally sharp edges of the laser pulse, Rabi oscillations described by Eq. (1) should be observed on the resonance wings. However, the presence of edges comparable to the duration of the pulse itself led to smoothing of these oscillations. To adequately describe this phenomenon, a trapezoidal dependence of the Rabi frequency of the step 1 on time was added to the theoretical model, as in Fig. 3. Only this made it possible to achieve good agreement between experiment and theory (green solid curves) in Fig. 5, and in Fig. 5(b) residual Rabi oscillations are visible on the wings both in the experiment and in the theory.

Figure 6(a) presents an experimental record (blue dots) of Rabi oscillations in three-photon laser excitation $5S_{1/2} \rightarrow 5P_{3/2} \rightarrow 6S_{1/2} \rightarrow 37P_{3/2}$ of the $37P_{3/2}$ Rydberg state at a power of the first-step laser radiation increased by ~10 times (compared to Fig. 4). For clarity, the experimental points are interpolated by spline function curves, since their number per oscillation is small due to the fact that the three-photon Rabi frequency increased to $\Omega/(2\pi) \approx 4.45$ MHz. In this case, the first oscillation with amplitude $\rho_{44} \approx 1 - \rho_{11} \approx 0.65$ occurs at a duration of 100 ns, which exactly corresponds to the spectrum in Fig. 5(a). A decrease in the resonance amplitude compared to Fig. 5(a) may be associated with insufficient time resolution in Fig. 6(a), which is limited by the electronics having a minimum time scanning step of 50 ns.

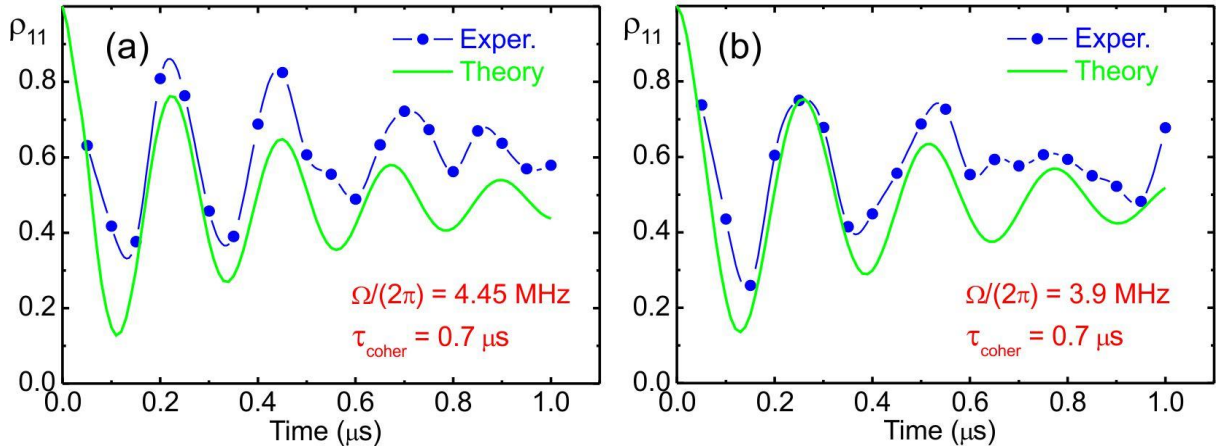

**Fig. 6.** Experimental records (blue dots) of the Rabi oscillations at three-photon laser excitation $5S_{1/2} \rightarrow 5P_{3/2} \rightarrow 6S_{1/2} \rightarrow 37P_{3/2}$ of the Rydberg state $37P_{3/2}$ at increased by ~10 times (compared to Fig. 4) power of radiation of the first-step laser. Green solid curves are the results of numerical modeling in a foul-level system with relaxation at one-photon Rabi frequencies $\Omega_1/(2\pi) \approx 260$ MHz, $\Omega_2/(2\pi) \approx 1740$ MHz $\Omega_3/(2\pi) \approx 30$ MHz. (b) The same at increased by 1.3 times power of the second-step laser, which increaser the Rabi frequency on the second step to $\Omega_2/(2\pi) \approx 2$ GHz

The green solid curve in Fig. 6(a) is the result of numerical simulation in the four-level model with relaxation at single-photon Rabi frequencies of $\Omega_1/(2\pi) \approx 260$ MHz, $\Omega_2/(2\pi) \approx 1740$ MHz, $\Omega_3/(2\pi) \approx 30$ MHz and taking into account the parasitic Zeeman broadening of the three-photon resonance. It can be noted that the theoretical curve reproduces well the amplitude and frequency of the observed Rabi oscillations, but does not accurately describe the stationary level, against which the damped Rabi oscillations occur. In the experiment, it is equal to $\rho_{44} \approx 1 - \rho_{11} \approx 0.4$, while the numerical calculation and analytical Eq. (4) predict that it should be equal to 0.5. We explain this phenomenon by two factors.

First, the detection of the Rydberg atom occurs with a delay of several microseconds relative to the end of the 1st step laser pulse. During this time, the population of the Rydberg state $37P_{3/2}$, which has an effective lifetime of 43 μs at room temperature, partially decays into the ground state. In the future, we plan to conduct experiments with longer-lived high states, where this effect will be suppressed.

Second, during the laser pulse, not only the Rydberg state is populated, but also the intermediate excited states $5P_{3/2}$ and $6S_{1/2}$ of the three-photon transition. In theoretical calculations, their total population is about 3%. At the end of the laser pulse, these states quickly decay into the ground state, which makes an additional contribution to the decrease in the stationary level of damped Rabi oscillations. To reduce this phenomenon, it is necessary to further increase the detuning of the intermediate states from resonances with laser radiation. This can be achieved by increasing the intensity of the 2nd step laser and the corresponding increase in the light shifts induced by it. However, in this case, the three-photon Rabi frequency will decrease.

To test the last statement, we increased the power of the second-step laser by 1.3 times, which increased the single-photon Rabi frequency of the second step to $\Omega_2/(2\pi) \approx 2$ GHz. The corresponding experimental record is shown in Fig. 6(b). It can be seen that this indeed led to a decrease in the three-photon Rabi frequency to $\Omega/(2\pi) \approx 3.9$ MHz. The observed change is also confirmed by theoretical calculation (green solid curve). Thus, by varying the power of the second step, it is possible to control the parameters of three-photon laser excitation of Rydberg states over a wide range. This significantly distinguishes three-photon excitation from the commonly used two-photon schemes, providing additional opportunities for increasing the contrast of Rabi oscillations and, ultimately, increasing the fidelity of quantum gates with Rydberg atoms.

## 6. Conclusion

In this paper, we present the results of our new experiment on three-photon laser excitation of a single [87]Rb Rydberg atom captured in an optical dipole trap. By narrowing the line width of the second and third excitation steps lasers, three-photon oscillations of the Rabi populations were observed for the first time with frequencies from 1 to 5 MHz, depending on the intensity of the laser pulses of the first and second excitation steps, at a coherence time of 0.7-0.8 μs. A specific feature of the experiment was the use of intense laser radiation with a wavelength of 1367 nm at the second excitation step, providing a single-photon Rabi frequency of up to 2 GHz to control the effective detunings of the intermediate levels of the three-photon transition due to the dynamic Stark effect.

The experiments revealed the presence of parasitic Zeeman broadening of the three-photon resonance from an insufficiently precisely compensated laboratory magnetic field in the region of a single atom. This led to a decrease in the contrast and coherence time of Rabi oscillations. To eliminate this effect, it is planned to implement optical pumping of a single atom to a certain Zeeman sublevel of the ground state and add a homogeneius magnetic field to create a Zeeman splitting of levels and excite three-photon transitions between specified Zeeman sublevels of the ground and Rydberg states.

It is also necessary to investigate the influence of the residual Doppler effect on the contrast and coherence time of Rabi oscillations of a single [87]Rb Rydberg atom captured in an optical dipole trap. Preliminary calculations have shown that even at the existing atomic temperature, estimated as 50-100 μK, the Doppler effect can be significantly suppressed by using counter-propagating laser beams of the first and the third excitation steps, in contrast to the co-propagating beams in the present experiment. Note that in a recent experiment on three-photon excitation of Rydberg states in an optical cell with thermal Cs atoms, due to such a configuration of laser beams, it was possible to almost completely suppress the Doppler effect and obtain an extremely narrow resonance of electromagnetically induced transparency with a width of 190 kHz [21].

All the above measures should allow us to significantly improve the contrast and coherence time of three-photon Rabi oscillations and, ultimately, increase the fidelity of quantum gates with Rydberg atoms compared to the fidelities achieved with two-photon laser excitation of Rydberg atoms.

This work was supported by the Russian Science Foundation (Grant. No. 23-12-00067, https://rscf.ru/project/23-12-00067/).